\documentclass[12pt,preprint]{aastex} 

\usepackage{graphicx}
\usepackage{amsmath}

\newcommand{\ec}{$\eta$~Car}

\begin{document}

\title{N II $\lambda\lambda$5668--5712, a New Class of Spectral Features in Eta Carinae\altaffilmark{1,2}}

\author{Andrea Mehner\altaffilmark{3}, 
        Kris Davidson\altaffilmark{3}, 
        Gary J.\ Ferland\altaffilmark{4}}

  \altaffiltext{1} {Based on observations made with the NASA/ESA Hubble Space Telescope. STScI is operated by the association of Universities for Research in Astronomy, Inc. under the NASA contract  NAS 5-26555.} 
  \altaffiltext{2} {Based on observations obtained at the Gemini Observatory, which is operated by the
Association of Universities for Research in Astronomy, Inc., under a cooperative agreement
with the NSF on behalf of the Gemini partnership.}
  \altaffiltext{3} {Department of Astronomy, University of Minnesota, 
       Minneapolis, MN 55455}   
  \altaffiltext{4} {Department of Physics \& Astronomy, University of 
       Kentucky, Lexington, KY 40506}  


\begin{abstract}

   We report on the \ion{N}{2} $\lambda\lambda$5668--5712 emission and 
   absorption lines in the spectrum of $\eta$ Carinae.        
   Spectral lines of the stellar wind regions can be classified into   
   four physically distinct categories:                     
   1) low-excitation emission such as \ion{H}{1} and \ion{Fe}{2}, 
   2) higher excitation \ion{He}{1} features,
   3) the \ion{N}{2} lines discussed in this paper, and 
   4) \ion{He}{2} emission. 
   These categories have different combinations of radial velocity behavior,
   excitation processes, and dependences on the secondary star. 
   The \ion{N}{2} lines are the only known features that originate 
      in ``normal'' undisturbed zones of the primary wind but depend 
      primarily on the location of the hot secondary star.       
   \ion{N}{2} probably excludes some proposed models, such
   as those where \ion{He}{1} lines originate in the secondary star's
   wind or in an accretion disk.             
  
\end{abstract}

\keywords{stars: emission-line, Be -
          stars: individual (eta Carinae) - stars: variables: general
          - stars: winds, outflows}

\section{Introduction}     

In $\eta$ Car's broad-line stellar wind spectrum, the high-excitation  
helium features have different profiles and fluctuate differently from 
the ``normal'' lines of \ion{H}{1}, \ion{Fe}{2}, etc.  Most authors now 
assume that the \ion{He}{1} emission depends on photoionization by  
a hot companion star, see   {\S}6 of  \citet{2008AJ....135.1249H} and 
references therein.   
Observed \ion{He}{2} emission is most likely excited by soft X-rays 
from unstable shocks
(\citealt{2006ApJ...640..474M}, but cf.\ \citealt{2004ApJ...612L.133S},     
\citealt{2007NewA...12..590K}, and \citealt{2006ApJ...652.1563S}).    
Features of these types are important because they        
    show direct influences by the secondary star and the           
    wind-wind collision region.                              
The only clear examples have been helium lines, whose source geometry  
is both complex and controversial.   In this paper we report similar 
characteristics in a set of \ion{N}{2} features, which  sample 
lower-ionization gas than the helium lines.

   Broad \ion{H}{1} and \ion{Fe}{2} emission lines represent     
   $\eta$ Car's stellar wind spectrum that would be present         
   even if the companion star did not exist \citep{2001ApJ...553..837H}.  Their main          
   components always remain close to system velocity
   (roughly $-8$ km s$^{-1}$, \citealt{1997AJ....113..335D,2004MNRAS.351L..15S}).

   The \ion{He}{1} emission lines, by contrast, shift progressively blueward  
   through most of the 5.5-year spectroscopic cycle.             
   When a  spectroscopic event occurs they show a more abrupt    
   negative shift, followed by a rapid positive reversal to       
   renew the cycle (see mid panel                                  
   of Figure \ref{fig:fig1}, and  figures in \citealt{2007ApJ...660..669N}).    
   The overall range of variation is about $0$                   
   to $-250$ km s$^{-1}$.   Evidently these changes represent         
   flows of highly ionized material, modulated by the secondary    
   star. As the secondary moves in its orbit, it progressively illuminates 
   regions with differing velocity fields, and may perturb some of them. 
   The details are model-dependent:                      
   compare, e.g., \citealt{2007ApJ...660..669N,2008MNRAS.386.2330D,2008AJ....135.1249H,2007NewA...12..590K,2008MNRAS.390.1751K,2006ApJ...640..474M,2001ASPC..233..173D}.                      
  
   \ion{He}{1} {\it absorption\/} shows a similar pattern with               
   variations between $-300$ km s$^{-1}$ and  $-550$ km s$^{-1}$.    
   This behavior is mirrored by the puzzling  \ion{He}{2} $\lambda$4687 
   emission observed during spectroscopic events 
   \citep{2004ApJ...612L.133S,2006ApJ...640..474M}.
    Hydrogen and \ion{Fe}{2} lines have components that behave qualitatively 
   like \ion{He}{1}, but cannot be measured separately 
   because they overlap the steady ``normal''  emission profiles.\footnote{
     This is particularly clear in STIS data showing H$\delta$ (Fig.\ 5 in 
     \citealt{2007ApJ...660..669N}).   Velocities quoted above are unpublished 
     measurements  of {\it Gemini\/}  GMOS observations during the 2009 
     spectroscopic event, see also \citet{2007ApJ...660..669N} and  
     \citet{2004ApJ...612L.133S} for similar values during the  
     2003.5 event.  }

Until now no other species were found in \ec\ with velocity 
variations like the helium lines, and without conspicuous steady components.  
Here we report that lines of the \ion{N}{2} 
$\lambda\lambda$5668--5712 multiplet 
do match this 
prescription. They are always present but were not discussed 
previously because they were weak.   
They are noteworthy because they depend on             
  a form of excitation by the hot secondary star, but in           
  different regions than the \ion{He}{1} lines.      
  They seem reasonable in some models, but are difficult to explain in others.       
For example, models where the helium emission originates in a small region close to the secondary star have difficulties in this respect; see {\S}4 below.                        

  Historically, hot windy objects such as P Cygni show strong \ion{N}{2} 
  emission and absorption (e.g., \citealt{1935MNRAS..95..580B,1935ApJ....81...66S,1940ApJ....91..546S}). Since nitrogen 
  is abundant in $\eta$ Car's CNO-processed wind, \citet{2001ApJ...553..837H} expressed surprise that \ion{N}{2} is not more  
  conspicuous there.  Some of the features discussed below were identified 
  by \citet{1953ApJ...118..234G}, but were too weak to be noted by \citet{1953MNRAS.113..211T}. More recently they have been detected by \citet{2001PhDT.........1Z}, \citet{1998A&AS..133..299D}, and \citet{2009yCat..21810473N} without discussion. The lines discussed here are physically different from [\ion{N}{2}] $\lambda$5756 which has 
 much lower energy levels.

\section{Data and Analysis}      

Ground-based slit spectroscopy of $\eta$ Car was obtained  with the 
{\it Gemini\/} Multi-Object Spectrograph (GMOS) on the Gemini South telescope 
from 2007 June to 2010 January.  
Here we only use 20 sets of observations with the B1200 line grating and a 0.5$\arcsec$  
slit that cover the spectrum at  $\lambda$5700 \AA\ on the star and on 
FOS4.\footnote{
   For a list of these observations and other details see the HST Treasury Program website http://etacar.umn.edu/,
   in particular Technical memo number 14.} 
FOS4 is a position in the SE lobe of the Homunculus (about 4{\arcsec} southeast of the star) which 
reflects the star's polar spectrum 
while the spectrum in direct view is representative of 
 lower latitudes      
\citep{1995AJ....109.1784D,
1999A&A...344..211Z,2001AJ....121.1569D,2003ApJ...586..432S}.
The spectral resolution was about 1.2 {\AA} or 65 km s$^{-1}$.
  The seeing varied 
from 0.5{\arcsec} to 1.5{\arcsec}, so each GMOS spectrum   
represents a region typically $\sim$ 1{\arcsec} across.
We prepared 2-D spectrograms with the standard GMOS data reduction pipeline 
in the Gemini IRAF package and extracted 1-D spectra.     The pipeline wavelength 
calibration was improved using the interstellar absorption line at 
$\lambda$5782 \AA;  
 in this paper we use heliocentric vacuum velocities.   
 Further details, unimportant for this paper, will be listed    
 in a later publication (Mehner et al.\ 2011, in preparation).

Eta Car was observed with the Hubble Space Telescope Space Telescope Imaging 
Spectrograph ({\it HST\/} STIS) from 1998 to 2004 and then again starting in 
2009. Here we only use  
observations with the G750M grating, the 52{\arcsec}$\times$0.1{\arcsec} 
slit, and central wavelength $\lambda$5734 \AA. 
The spectral resolution was about 1.1 {\AA} or 60 km s$^{-1}$.
The 1998--2004 observations  
include  a variety of slit positions and orientations. 2009 June and December 
data cover a region up to 1{\arcsec} from the central source with parallel 
slit positions offsets of 0.1{\arcsec}. 
    The long-exposure 2009 data were obtained in {\it HST\/} programs 
    11506 and 12013, whose PI's were K.\ S.\ Noll and M.\ Corcoran. 
   The data were reduced using 
routines developed by the HST Treasury Project that include several 
improvements over the normal STScI pipeline and standard CALSTIS 
reductions.\footnote{
    For information see http://etacar.umn.edu and \cite{2006hstc.conf..247D}.} 
We extracted spectra 0.1{\arcsec} wide.

\section{Results}
\label{sec:results}

\begin{figure}   
  \epsscale{0.5}
  \plotone{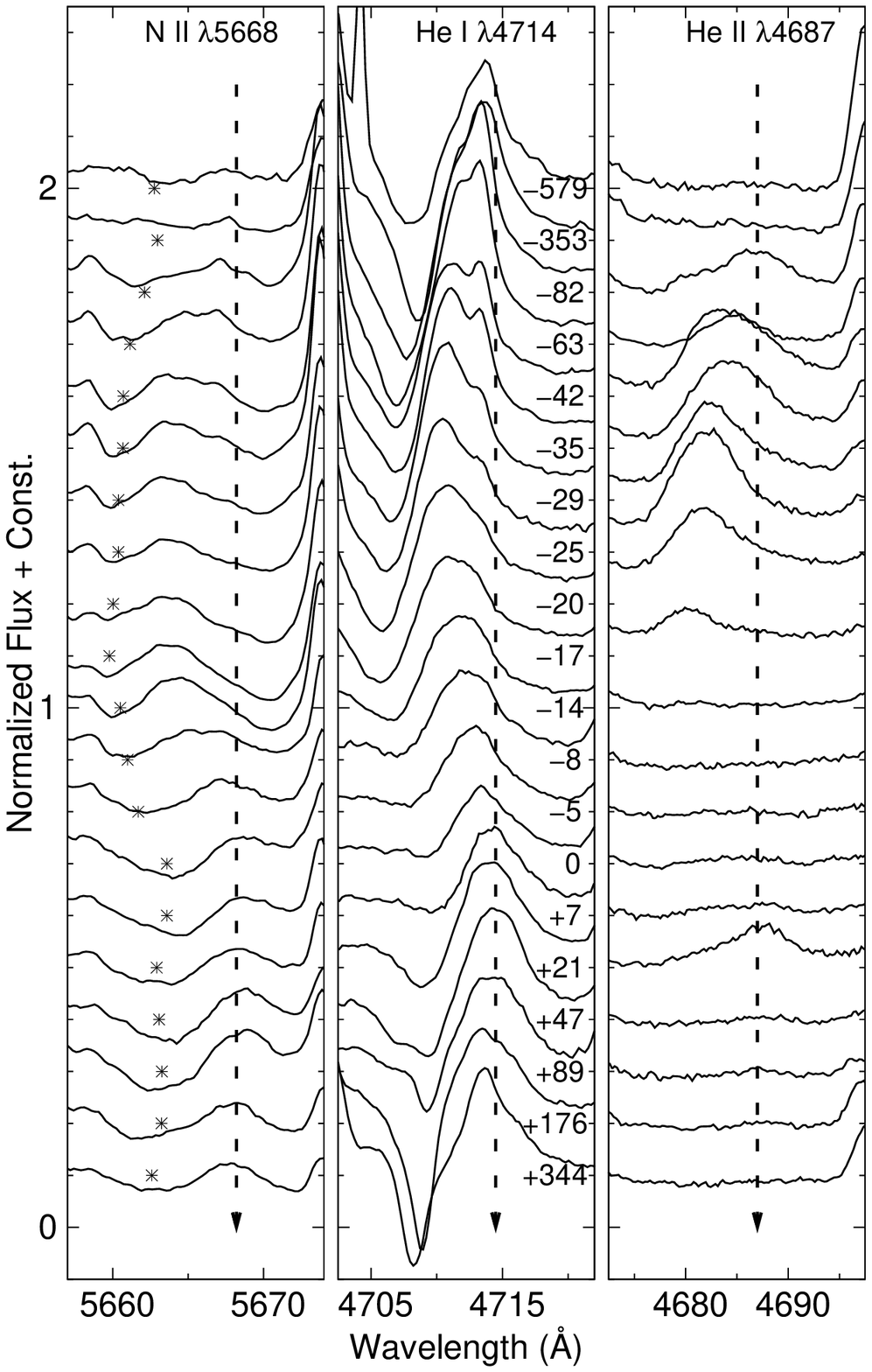}
  \caption{\ion{N}{2} $\lambda$5668, \ion{He}{1} $\lambda4714$, and \ion{He}{2} $\lambda4687$ in spectra
 of \ec\ obtained with {\it Gemini\/} GMOS from 2007 June to 2010 January. Days before and after the 2009
 spectroscopic event at MJD 54860 are displayed next to each tracing. Arrows indicate the zero velocities
 of the emission lines.  {\it Note that most of the 5.5 year cycle is represented by the top two and bottom two samples and that  [\ion{Fe}{2}] $\lambda$5675 blends with the \ion{N}{2} $\lambda\lambda$5678,5681 absorption.}        
  \label{fig:fig1}}
  \end{figure}

\begin{figure}   
  \epsscale{0.5}
  \plotone{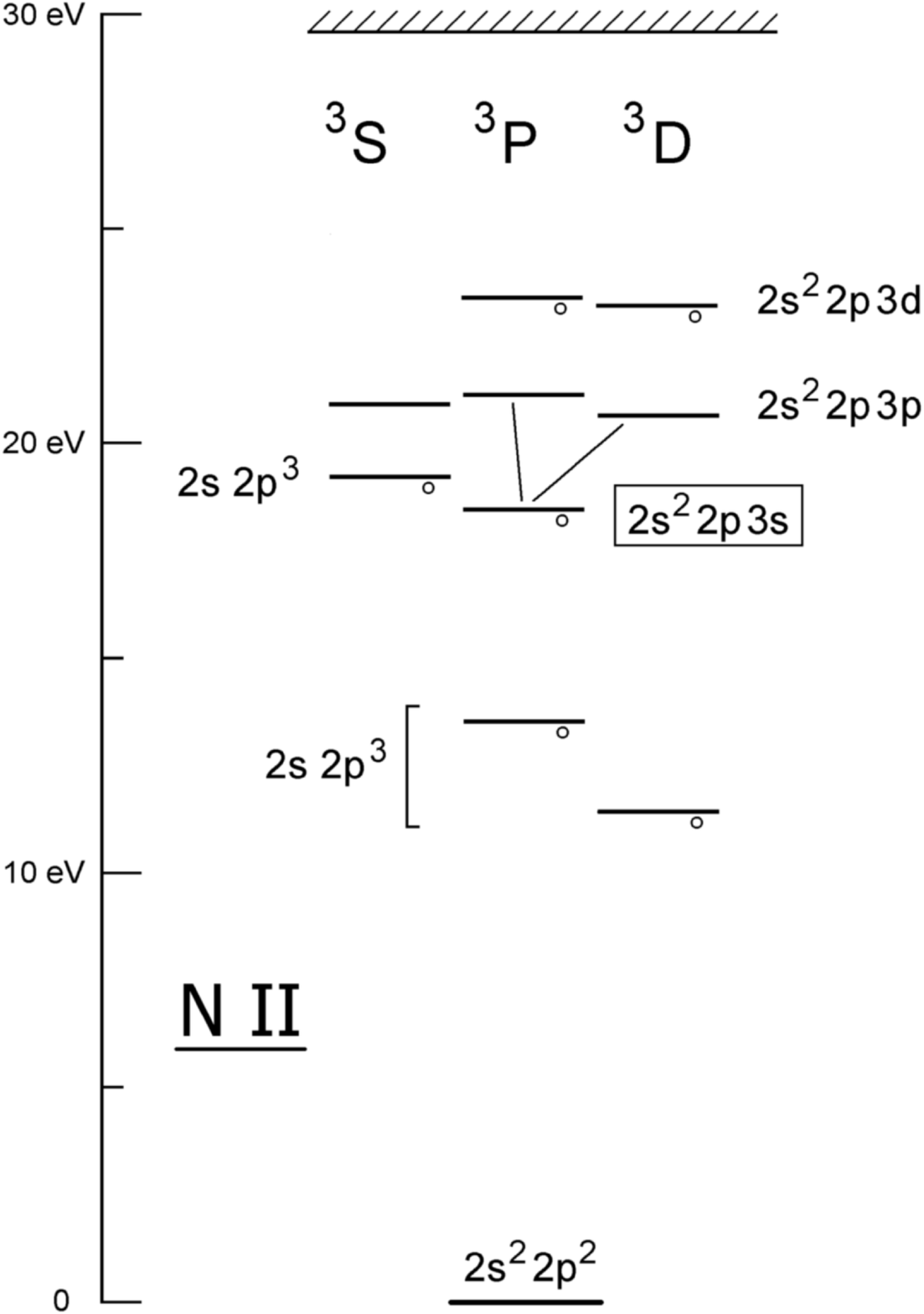}
  \caption{The lowest triplet levels of \ion{N}{2}.  Odd-parity 
     levels such as $2s^2 2p 3s$ $^3$P$^\mathrm{o}$ are marked with 
     small `o's.  In this paper we concentrate on the 
     $2s^2 2p 3s$ -- $2s^2 2p 3p$ transitions because they  
     have observationally convenient wavelengths, see {\S}3.          
  \label{fig:fig2}}
  \end{figure}

In the GMOS data we found  
broad emission and absorption lines of the \ion{N}{2} 
$2s^{2} 2p 3s$ $^3$P${^\mathrm{o}}$ -- $2s^{2} 2p 3p$ $^3$D multiplet  
at $\lambda\lambda$5668--5712 \AA,  
 exhibiting       
radial velocity variations during the 2009 spectroscopic event similar to the 
helium lines;  see Figure \ref{fig:fig1} and Table \ref{tab:table1}.  
Figure \ref{fig:fig2} shows the energy levels involved.
Note that the lower level, $2p 3s$ $^3$P$^\mathrm{o}$,  is more than 18 eV 
above the \ion{N}{2} ground level but is connected to it by a permitted 
transition.  We discuss the excitation mechanism below.

 Figure \ref{fig:fig1} compares 
spectra of \ec\ obtained from 2007 June to 2010 January 
showing \ion{N}{2} $\lambda$5668, \ion{He}{1} $\lambda4714$, and  
\ion{He}{2} $\lambda4687$. Weak \ion{N}{2} $\lambda\lambda$5668--5712 emission and 
absorption is always present in the spectrum of \ec. The maximum strength 
of the absorption component has a radial velocity of about 
$-300$ km s$^{-1}$ with respect to the emission 
peak.    
During the 2009 spectroscopic event, emission and absorption features 
shifted about 
$250$ km s$^{-1}$ towards the blue, simultaneous with the \ion{He}{1} 
emission and absorption lines and the \ion{He}{2} $\lambda4687$ emission. 

{\it HST\/} STIS data obtained in 2009 June and December clearly show the \ion{N}{2} $\lambda\lambda$5668--5712
 lines, see Figure \ref{fig:fig3}.  
 In retrospect the faint \ion{N}{2} spectral features can also
 be detected in {\it HST\/} STIS data from 1998--2004,  but due to lower S/N 
 they failed to attract notice before.
Like the \ion{He}{1} absorption, \ion{N}{2} absorption was much weaker before 2003; less than 30\% compared to 2009 (compare \citealt{2010ApJ...717L..22M}). 
The absorption strengths of both species increased gradually over the last 10 years except during the 2003.5 and 2009.0 spectroscopic events.
 We determined the ``stellar continuum'' by a loess (locally weighted scatterplot
 smoothing) fit to the spectrum and find that the \ion{N}{2} absorption is stronger than the emission.
        (The same is true for P Cygni; \citealt{1935MNRAS..95..580B}, \citealt{1935ApJ....81...66S}).
 {\it Gemini\/} GMOS data appear to indicate the opposite, but their data quality is too low to determine a
 reliable continuum, and ground-based 
spectra are contaminated by narrow emission lines from ejecta outside the 
wind, e.g., [\ion{Fe}{2}]
 $\lambda$5675 blends with the \ion{N}{2} $\lambda\lambda$5678,5681 absorption. 
Although we first noticed these features in GMOS  data, the high 
spatial resolution of {\it HST\/} data is essential for examining their character.   

Most other permitted \ion{N}{2} lines 
are too weak or blended with emission lines from the wind or nearby ejecta 
to be observed.  However, we see similar behavior in the \ion{N}{2} $2p 3s$ $^3$P${^\mathrm{o}}$ -- $2p 3p$ $^3$P  series at $\lambda\lambda$4603,4608 \AA\ (Fig.\ \ref{fig:fig2}). Transitions of the  \ion{N}{2} $2p 3s$ $^3$P${^\mathrm{o}}$ -- $2p 3p$ $^3$S  series are too weak and blended with other lines.  

Radiative excitation of level $2s^2 2p 3s$ is very important as discussed later. The $2s 2p^3$ levels are excited in the same way but lead to no observable results. Transitions from $2s 2p^3$ $^3$S$^\mathrm{o}$ to $2s^2 2p 3p$ ($\lambda\lambda$6435--7265 \AA) have small oscillator strengths and are therefore not observed.
STIS data indicates the weak presence of the singlet \ion{N}{2} $\lambda$3996 line but this line is blended with others in {\it Gemini\/} data and therefore the identification is not certain. Lines whose upper levels are above the $2p 3p$ state like \ion{N}{2} $\lambda$5007, even though they are strong in the laboratory, are not observed. 

The \ion{N}{2} lines and their velocity shifts 
can also be seen in the reflected polar spectrum at location FOS4, the location decribed in {\S}2. The velocity shifts from the FOS4 separate direction may cast doubt on orbit models, but this question is too complex to discuss here.

\section{Line Formation and Significance}    

The same \ion{N}{2} lines are seen in early type stars. 
In objects like P Cygni and shells of O-stars, 
   $2p3s$ $^1$P$^\mathrm{o}$ -- $2p3p$ $^1$S,$^1$P,$^1$D and 
   $2p3s$ $^3$P$^\mathrm{o}$ -- $2p3p$ $^3$S,$^3$P,$^3$D are very strong 
\citep{1940ApJ....91..546S}. Large differences
 between WN stars indicate that these lines are sensitive to atmospheric conditions and/or the variability
 of the wind. They may also be very sensitive to the FUV flux since they are likely produced via continuum
 fluorescence \citep{2001ApJ...548..932H}.  

In the case of $\eta$ Car the \ion{N}{2} absorption and emission almost 
certainly occur in the primary star's dense wind, for reasons noted 
later below.  But three facts indicate that UV photons from the secondary 
star populate the critical $2p3s$ level.        
(1) The velocity behavior described in {\S}3 strongly suggests some 
relation to the secondary,  analogous to the \ion{He}{1} 
features.  (2) The $2p3s$ levels are about 18.5 eV above the ground 
state, a high value for the primary star's wind. 
According to \citet{2001ApJ...553..837H}, the opaque-wind photospheric  
temperature is below $15,000$ K and emission-line regions in the 
primary wind are mostly below $10,000$ K;  much cooler than the 
O stars and WR objects mentioned above.  The hot secondary star, on 
the other hand, produces a large flux of 18.5 eV photons 
(Fig.\ 10 in \citealt{2010ApJ...710..729M}).  
(3) This hypothesis appears quantitatively 
successful as outlined below.  Any of these clues by itself might 
be debatable, but together they seem compelling to us.  

Let us summarize an order-of-magnitude assessment of the absorption 
line strengths that one would expect in a simple model.          
For simplicity we include only the N$^+$ ground level `1' ($2p^2$ $^3$P) 
and one excited level `2' ($2p 3s$ $^3$P$^\mathrm{o}$);  our initial 
goal is to estimate the equilibrium population ratio  $n_2/n_1$.
A two-level system is justified because no permitted transitions
connect level 2 to intermediate levels.
Consider a uniformly-expanding local volume in the primary wind.  Denote 
the incident continuum photon flux at $h{\nu}_{12} \approx 18.5$ eV 
by ${\Phi}_{\nu}$, not corrected for line absorption in the gas.
As explained in {\S}8.5 in \citealt{1999isw..book.....L}  
and {\S}8 in \citealt{2006ApJ...640..474M}, the Sobolev approximation 
allows us to write expressions for the radiative $1 \rightarrow 2$ 
excitation rate and the re-emission photon escape probability, as 
functions of the local expansion rate, a line profile function, and 
atomic parameters.
The {\it effective\/} de-excitation rate is proportional to the escape 
probability, and the most complicated factors appear similarly 
in both the excitation rate and the escape probability. Consequently 
the equilibrium population ratio is simple:    
    \begin{equation}  
    \frac{n_2}{n_1}  =  
        \frac{g_2}{g_1} \cdot \frac{ {\lambda}_{12}^2 {\Phi}_{\nu} }{8 \pi} \; , 
    \label{n2n1eqn} 
    \end{equation} 
where $g_1$ and $g_2$ are the statistical weights.  This expression 
remains valid when we take fine structure into account. 
The same ratio would be found in an LTE case where 
${\Phi}_{\nu}/c$ is the Planckian photon density at 
$h{\nu}_{12} \approx 18.5$ eV.   Above we mentioned triplet levels
of \ion{N}{2}, but singlet levels would also be excited in the same 
way from $2p^2$ $^1$D.

With conventional parameters for the two stars and their winds, 
${\Phi}_{\nu}$ at 18.5 eV is dominated 
by the hot secondary star.  Reasonable values for it are   
$T_\mathrm{eff} \approx 40,000$ K, $R \approx 13 \; R_{\odot}$ and $L \approx 4 \times 10^5 \; L_{\odot}$  
\citep{2010ApJ...710..729M}.  According to a WM-basic atmosphere model \citep{2001A&A...375..161P}, such  
an object emits roughly $5 \times 10^{33}$ photons Hz$^{-1}$ s$^{-1}$  
at $h{\nu} = 18.5$ eV, about 30\% less than a $40,000$ K blackbody.  
Figure  \ref{fig:fig4} shows the probable geometry.
With the type of orbit model that most authors currently favor 
   (e.g., \citealt{2008MNRAS.388L..39O}),
the separation between stars was about 13 AU when $\eta$ Car was observed 
in 2009 June.  The secondary star was then located roughly 1--5 AU closer 
to us than the primary, and roughly 10 AU from our line of sight to 
the primary -- depending  on the poorly known orbit orientation, 
of  course.  With these parameters, eqn.\ 1 gives 
$n_2/n_1 \approx 3 \times 10^{-8}$ at relevant locations between us 
and the primary star, i.e., in gas located about 10 AU from the secondary 
star. The equivalent excitation temperature is near $12,500$ K.

The observed absorption lines have some geometrical complications.
Since the opaque-wind photosphere is diffuse with a substantial size,  
a relevant ``line of sight'' 
can be any sample ray in a cylinder with diameter $\sim$ 7 AU (Fig.\ 
\ref{fig:fig4}).   Strong $2p3s$--$2p3p$ absorption occurs in regions  
that optimize a combination of attributes:   (1) the inner wind is 
favored because of its high densities; (2) most of the nitrogen must 
be singly ionized;  (3) the N$^+$  must be fairly close to the secondary 
star in order to maximize $\Phi_{\nu}$; and (4) the light path (possibly 
indirect, due to Thomson scattering) must pass close enough to the 
primary star to sample continuum radiation from its semi-opaque inner wind.  
Thus, in Fig.\ \ref{fig:fig4}, we expect the strongest absorption to 
occur near and above the symbol `1' that labels Region 1.  The picture 
obviously changes as the secondary  star moves along its orbit.  

Most nitrogen in the primary wind is singly ionized, due to both the 
primary and the secondary radiation field \citep{2001ApJ...553..837H}.   
Helium in zones 3 and 4 of Fig.\ \ref{fig:fig4} provides the shielding which prevents the secondary radiation field
from ionizing N$^{+}$ into N$^{++}$, but which allows the 18.5 eV radiation to penetrate.
Suppose the primary mass-loss rate is 
$3 \times 10^{-4}$ $M_{\odot}$ yr$^{-1}$ and nearly all the CNO is 
nitrogen \citep{1986ApJ...305..867D,1999ASPC..179..134D}.\footnote{                            
    The most often quoted mass loss rate for $\eta$ Car is around  
    $10^{-3}$ $M_\odot$ yr$^{-1}$, but there are good reasons to 
    believe that it has decreased substantially in the past decade 
    \citep{2010ApJ...717L..22M,2010ApJ...725.1528C,2009ApJ...701L..59K}.
    }  
Then the column densities outside $r = 4$ AU (for example) are  
  \begin{displaymath}  
  N(\mathrm{N}^{+}) \; \sim \; 2 \times 10^{20} \; \mathrm{cm}^{-2} \, , 
  \end{displaymath} 
  \begin{displaymath}
  N(\mathrm{N}^{+} \, 2p3s) \; \sim \; 6 \times 10^{12} \; \mathrm{cm}^{-2} \, .  
  \end{displaymath}
This column density would produce a combined equivalent width 
of about 1 {\AA} for the $\lambda$5681.14 and $\lambda$5677.60 \AA\ absorption 
lines. Since this exceeds the observed value of 0.5 {\AA}, 
{\it the proposed mechanism appears to be sufficient,\/}  
even if only a limited part of the wind is involved.  
Let us emphasize that the parameter values 
in this sample calculation were chosen {\it ab initio\/} without knowing 
what result they would lead to;  there was no readjustment 
to get a desired outcome. Note that the margin of error does not allow much smaller values of ${\Phi}_{\nu}$ and gas densities.

The {\it emission\/} features reported in {\S}3 are comparable in strength 
to the absorption -- probably weaker and certainly not much stronger -- 
so they resemble the pure scattering case of P Cyg lines. 
In other words, the observed  $2p3s$ -- $2p3p$ emission line  
strengths automatically resemble the absorption strengths.  
More detailed calculations will require elaborate geometrical models of 
the ionization and excitation zones.  
  
As we noted earlier, these \ion{N}{2} features should arise chiefly 
in regions of the primary wind that are fairly close to the secondary 
star, and, therefore, close to the He$^+$ zones and wind-wind shocks.  
If shock instabilities do not distort it too much, the He$^{+}$/He$^{0}$ 
ionization front is expected to have a pseudo-hyperboloidal shape that 
may be either concave or convex towards the secondary star (Fig.\ \ref{fig:fig4}).  
Adjoining primary-wind zones are  spatially large enough to account for the 
\ion{N}{2} lines.  Generically, this type of model can explain the velocity 
variations, but authors disagree about details (e.g., \citealt{2007ApJ...660..669N,2008MNRAS.386.2330D,2008AJ....135.1249H,2006ApJ...640..474M}).

Far more important than merely being consistent with some models 
of the $\eta$ Car system,  the \ion{N}{2} features may disprove 
others.   For example, it has been suggested that the helium lines originate 
in the acceleration zone of the secondary star's wind, or perhaps in an 
accretion disk around the secondary, rather than in the primary wind  
  \citep{2006ApJ...652.1563S,2007NewA...12..590K,2004ApJ...612L.133S}.   
But the \ion{N}{2} features, with practically the same velocity behavior, 
almost certainly cannot represent such zones;  so models of that type 
would be forced to explain the \ion{He}{1} and \ion{N}{2} velocities 
differently.  

\begin{figure}   
  \epsscale{0.5}
  \plotone{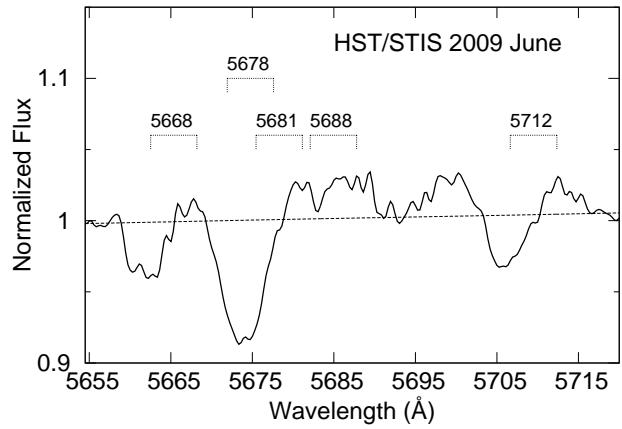}
  \caption{\ion{N}{2} $\lambda\lambda$5668--5712 in {\it HST\/} STIS data from 2009 June. Zero velocity emission
 and absorption at $-300$ km s$^{-1}$ are indicated with brackets. The dashed curve represents the stellar
 continuum. {\it HST's\/} high-spatial resolution shows the absorption much better than ground-based data. The strong \ion{N}{2} $\lambda\lambda$5678,5681 absorption is blended with [\ion{Fe}{2}] $\lambda$5675 in the {\it Gemini\/} spectra.
  \label{fig:fig3}}
  \end{figure}

\begin{figure}   
  \epsscale{0.5}
  \plotone{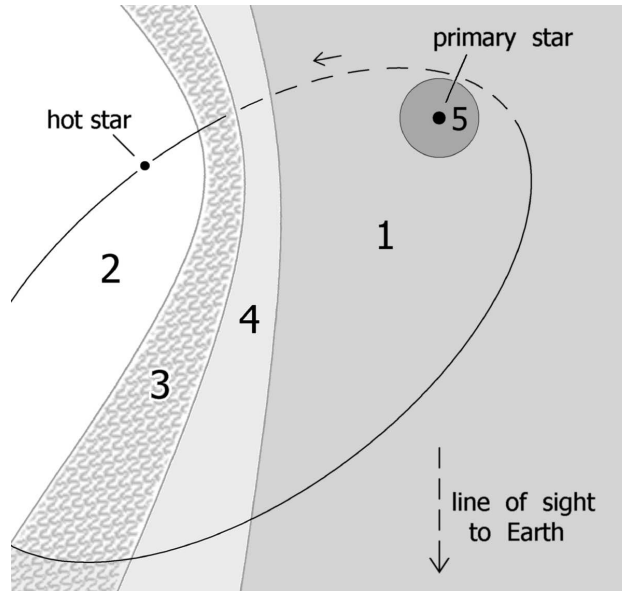}
  \caption{ Schematic arrangement of a conventional model for the 
    2009 June observations.  The line of sight to Earth lies in the plane of the Figure, and the 
    secondary star's orbit is shown projected onto that plane.   Regions 1 and 2 are undisturbed parts of the primary 
    and secondary winds.   The shocked wind-wind collision region is 
    labeled 3, while region 4 is a \ion{He}{2} zone in the primary 
    wind, photoionized by the secondary star.  \ion{N}{2} is abundant 
    in region 1 but zones 2--4 are more highly ionized.  Region 5 
    is within the opaque-wind photosphere at visual wavelengths.  
    The area shown is roughly 25 AU across, but all details are    
    simplified and idealized.
  \label{fig:fig4}}
  \end{figure}

The arguments are simplest   
for absorption lines.  First we note that with credible parameters  
for the two stars (e.g., \citealt{2001ApJ...553..837H,2010ApJ...710..729M}),
the primary wind accounts for at least 95\% of the visual-wavelength 
continuum, most likely 98--99\%.  Therefore {\it the absorption features 
in Fig.\  \ref{fig:fig3} definitely represent material located 
between us and the inner parts of the primary wind.\/}   
(In other words, blocking the entire secondary star would not produce 
absorption as deep as that shown in Fig.\ \ref{fig:fig3}, even if 
$\tau_\mathrm{line} \gg 1$.)  
Can this absorbing material be part of the secondary wind or an 
accretion disk of the secondary star?  Presumably not; because in 
every proposed orbit model the relevant lines of sight either miss the 
secondary star and its wind entirely at most times, or pass through 
only the outskirts of the secondary wind.  Gas densities in the secondary wind are 
two orders of magnitude less  than we used in the calculations outlined 
above, the nitrogen is mainly \ion{N}{3}, and the velocity dispersion 
is too large for the observed absorption features.  This type of model  
would badly fail the quantitative feasibility test that the  
``conventional'' model passed.  Regarding the \ion{N}{2} emission 
features, they are fairly weak relative to the total observed 
continuum, but would be extraordinarily strong compared to the 
visual-wavelength continuum of just the secondary star.

Developing these arguments further would be beyond the scope of this
paper;  our main point is that {\it the \ion{N}{2} lines convey
information that is not available in other spectral features.}
This is mainly because they arise in ``normal'' parts of the primary
wind but they are sensitive to the current location of the secondary
star.  Helium lines also depend on UV from the secondary, but their
emission zones are extremely difficult to model since they depend on
the unstable primary shock, post-shock cooling, local "clumping," etc.
In principle the N II zones are expected to be much steadier, and
the excitation mechanism is relatively insensitive to inhomogeneities.
For these reasons the \ion{N}{2} features may be good indicators of
average density in the wind combined with the secondary star's orbital
motion.  In particular one might expect their velocities to be easier
to model than those of helium lines.

If the ``conventional'' model is more or less valid, then the   
\ion{N}{2} absorption features should weaken around apastron when 
$\Phi_{\nu}$ is smallest.  Unfortunately the 2000--2002 {\it HST\/} STIS 
data (the only previous apastron passage when STIS was available) 
are not  adequate to test this ``prediction'' easily.  Perhaps 
new STIS observations will be made in 2011--2012.

In {\S}3 we mentioned that \ion{He}{1} and \ion{N}{2} absorption 
lines have grown in strength during the past 10 years.  At first sight 
this appears counter-intuitive, since the wind density appears 
to have {\it decreased\/} \citep{2010ApJ...717L..22M}.  
However, the situation has various dependencies that tend to oppose 
each other so intuition may be a poor guide.  Decreasing mass loss 
rate implies a smaller, denser opaque-wind photosphere;  eqn.\ 1 becomes 
invalid if continuum absorption destroys the trapped 18.5 eV resonance 
photons before they escape; etc.  This problem merits further 
attention. As the wind decreases it is conceivable that the primary star may eventually become hot enough to excite these lines in addition to the secondary.  Whatever the solution, we expect the \ion{N}{2} 
lines to provide different parameter contraints than the 
spectral features that are excited in more normal ways.

In summary, spectral features in $\eta$ Car's wind can be assigned to these 
categories:
\begin{enumerate} 
  \item The ``normal'' lines of hydrogen, \ion{Fe}{2}, etc.  In principle  
     these can be used to analyze the primary wind 
     \citep{2001ApJ...553..837H}.   
  \item \ion{He}{1} emission and absorption.  Most authors agree that these 
     are related in some way to the secondary star, but the details are 
     controversial (see references cited in {\S}1 above).  
  \item \ion{N}{2} emission and absorption.  These lines arise in normal-ionization parts of the primary  
       wind, but they depend mainly on the proximity of the secondary 
       star.  In this sense they sample a new region of parameter 
       space. In most recent models, the \ion{N}{2} zones are much simpler than the ionized helium.
       We emphasize that the \ion{He}{1} lines depend on photoionization,
$h\nu > 24.6$ eV, whereas the \ion{N}{2} lines depend on photoexcitation at 
$h\nu \sim 18.5$ eV. The ratio $f_{\nu}(> 25~\mathrm{eV})/f_{\nu}(18.5~\mathrm{eV})$ depends strongly on the secondary star's temperature (WM-basic atmosphere models, \citealt{2001A&A...375..161P}). 
  \item \ion{He}{2} $\lambda$4687, an unsolved puzzle not discussed here.  
  It depends on soft X-rays from the colliding-wind region, not UV from the secondary star.
\end{enumerate}  
A good basic model for the primary wind and the binary system must account 
for at least the first three of these, throughout most of the 5.5-year 
cycle.  (\ion{He}{2} $\lambda$4687 may depend on shock instabilities too complex 
for a ``basic'' model.)

After an earlier version of this paper appeared on astro-ph,
 \citet{2011arXiv1104.4655K} have proposed  
       an alternative interpretation of the \ion{N}{2} lines.  A full 
       response would make this paper too long, but in our view they  
       do not explain how gas in a very small locale near the secondary 
       star can produce substantial absorption lines in the spectrum of 
       the much larger primary-wind photosphere (see discussion above).

{\it Acknowledgements}
We thank the staff and observers of
the Gemini-South Observatory in La Serena for their help in preparing and conducting the observations.
We also thank Roberta Humphreys, John Martin, and Kazunori Ishibashi for discussions and helping with non-routine steps in the data preparation.


\begin{deluxetable}{lcccc}    
\tabletypesize{\scriptsize} 
\tablecaption{Observed \ion{N}{2} $\lambda\lambda$5668--5712 transitions\tablenotemark{a}\label{tab:table1}}   
\tablewidth{0pt} 
\tablehead{ 
\colhead{Wavelength} & 
\colhead{Transition} & 
\colhead{E$_{i}$} &
\colhead{E$_{k}$} &
\colhead{ A$_{ki}$}  \\
\colhead{(\AA)} & 
\colhead{} & 
\colhead{(cm$^{-1}$)} &
\colhead{(cm$^{-1}$)} &
\colhead{(s$^{-1}$)} 
 }   
\startdata
N II 5668.20 & $2s^2 \; 2p \; 3s$ $^{3}$P$_{1}$ -- $2s^2 \; 2p \; 3p$ $^{3}$D$_{2}$ & 148940.17  	&
 	 166582.45  & 3.45e+07 \\
N II 5677.60 & $2s^2 \; 2p \; 3s$ $^{3}$P$_{0}$ -- $2s^2 \; 2p \; 3p$ $^{3}$D$_{1}$ & 148908.59  	&
 	 166521.69  & 2.80e+07\\
N II 5681.14 & $2s^2 \; 2p \; 3s$ $^{3}$P$_{2}$ -- $2s^2 \; 2p \; 3p$ $^{3}$D$_{3}$ & 149076.52  	&
 	 166678.64  & 4.96e+07 \\
N II 5687.79 & $2s^2 \; 2p \; 3s$ $^{3}$P$_{1}$ -- $2s^2 \; 2p \; 3p$ $^{3}$D$_{1}$ & 148940.17  	&
 	 166521.69  & 1.78e+07 \\
N II 5712.35 & $2s^2 \; 2p \; 3s$ $^{3}$P$_{2}$ -- $2s^2 \; 2p \; 3p$ $^{3}$D$_{2}$ & 149076.52  	&
 	 166582.45  & 1.17e+07 \\
\enddata 
\tablenotetext{a}{http://physics.nist.gov/PhysRefData/ASD}
\end{deluxetable}  

\end{document}